\documentclass[showkeys,showpacs,superscriptaddress]{revtex4}
\usepackage{amsmath}
\usepackage{amssymb}
\usepackage{graphicx}

\DeclareMathAlphabet{\mathpzc}{OT1}{pzc}{m}{it}
\begin{document}
\title{
On the linear stability of polytropic fluid spheres in $R^2$ gravity
}

\author{
Vladimir Dzhunushaliev
}
\email{v.dzhunushaliev@gmail.com}
\affiliation{
Department of Theoretical and Nuclear Physics,  Al-Farabi Kazakh National University, Almaty 050040, Kazakhstan
}
\affiliation{
Institute of Experimental and Theoretical Physics,  Al-Farabi Kazakh National University, Almaty 050040, Kazakhstan
}
\affiliation{
Academician J.~Jeenbaev Institute of Physics of the NAS of the Kyrgyz Republic, 265 a, Chui Street, Bishkek 720071, Kyrgyzstan
}

\author{Vladimir Folomeev}
\email{vfolomeev@mail.ru}
\affiliation{
Institute of Experimental and Theoretical Physics,  Al-Farabi Kazakh National University, Almaty 050040, Kazakhstan
}
\affiliation{
Academician J.~Jeenbaev Institute of Physics of the NAS of the Kyrgyz Republic, 265 a, Chui Street, Bishkek 720071, Kyrgyzstan
}
\affiliation{
International Laboratory for Theoretical Cosmology, Tomsk State University of Control Systems and Radioelectronics (TUSUR),
Tomsk 634050, Russia
}

\begin{abstract}
Within $R^2$ gravity,
we study the linear stability of strongly gravitating spherically symmetric
configurations supported by a polytropic fluid. All calculations are carried out
in the Jordan frame. It is demonstrated that, as in general relativity,
the transition from stable to unstable systems occurs at the maximum of the curve mass-central density of the fluid.
\end{abstract}




\maketitle
\section{Introduction}

Einstein's general relativity (GR) is a generally recognized theory of gravitation that is successfully applied
on different spatial and time scales.  In particular, it is used to model processes which took place in the early Universe.
It is, however, obvious that to describe the very early stages of the evolution of the Universe, it is already needed to take into account quantum effects.
To do this, it is necessary to have a full theory of quantum gravity, which is absent at present. For this reason, starting from  1960's,
it was being suggested to use some effective description of quantum effects in strong gravitational fields by a change
of the classical Einstein gravitational Lagrangian $\sim R$ by various
modified Lagrangians containing different curvature invariants. In the simplest case this can be some function  $f(R)$ of the scalar curvature $R$.
Such modified gravity theories (MGTs) have been widely applied to
 model various cosmological aspects of the early and present Universe
(for a general review on the subject, see, e.g., Refs.~\cite{Nojiri:2010wj,Nojiri:2017ncd}).

On the other hand, in considering processes and objects on relatively small scales comparable to sizes of galaxies and even of stars,
the effects of modification of gravity can also play a significant role.
For example, within $f(R)$ gravity, there have been constructed models of
relativistic stars~\cite{rel_star_f_R_1,rel_star_f_R_2},
 wormholes~\cite{wh_f_R_1,wh_f_R_2},  and neutron stars~\cite{Cooney:2009rr,Arapoglu:2010rz,Orellana:2013gn,Alavirad:2013paa,Astashenok:2013vza,Ganguly:2013taa}.
It was shown that the modification of gravity can affect, in particular, a number of important physical characteristics of neutron stars which may be verified observationally.
One of the problems of interest is determining the mass-radius (or the mass-central density of matter) relations,
which have been obtained, for example, in Refs.~\cite{Astashenok:2014pua,Astashenok:2014dja,Capozziello:2015yza,Bakirova:2016ffk,Astashenok:2017dpo,Folomeev:2018ioy,Feola:2019zqg,Astashenok:2020isy}.
It was shown there that, as in GR, in MGT, for some central density of neutron matter, the
mass-central density curves possess maxima whose location depends on the physical properties of the specific matter.
In GR, the presence of such a maximum indicates the fact that there is a transition from stable systems (located to the left of the maximum) to unstable configurations
(located to the right of the maximum).
This fact is established by studying the behavior of matter and metric perturbations using the variational approach~\cite{Chandrasekhar:1964zz}.

In MGTs, different types of perturbations have been repeatedly studied as well (see, e.g., Refs.~\cite{Blazquez-Salcedo:2018pxo,Blazquez-Salcedo:2020ibb} and references therein).
In considering these problems, a transition from $f(R)$ gravity to a scalar-tensor theory is usually performed.
Correspondingly, studies of the perturbations are carried out not
in the Jordan frame but in the Einstein frame. However, the question of the physical equivalence between these two frames is still under discussion,
and it cannot be regarded as completely solved~\cite{Capozziello:2010sc,Kamenshchik:2014waa,Kamenshchik:2016gcy,Bahamonde:2016wmz,Ruf:2017xon}.
Here, the following potential difficulties may be noted:
(i)~ In performing the transition from $f(R)$ gravity to a scalar-tensor theory, there may, in general, occur some undesirable consequences (singularities, fixed points, etc.); as a
 result, the equivalence between the frames can be violated.
 (ii)~Objects that are stars in one frame may represent some other configurations in another.
(iii)~In constructing perturbation theory, the equivalence between the frames can be lost in view of the approximate nature of such a theory.

In this connection, it may be of some interest to study the stability directly in the Jordan frame, and this is the goal of the present paper.
For the sake of simplicity,  we will work within $R^2$ gravity, where linear radial perturbations of a strongly gravitating system supported by a polytropic fluid will be investigated.
For this purpose, in Sec.~\ref{gen_eqs}, we first derive the general equations for $f(R)$ gravity.
Using these equations, in Sec.~\ref{stat_conf}, we numerically find static solutions describing equilibrium configurations, on the background of which the behavior of matter and spacetime
perturbations is studied in Sec.~\ref{stab_anal}.

\section{General equations}
\label{gen_eqs}

We consider modified gravity
with the action [the metric signature is $(+,-,-,-)$]
\begin{equation}
\label{action_mod}
S=-\frac{c^3}{16\pi G}\int d^4 x \sqrt{-g} f(R) +S_m,
\end{equation}
where $G$ is the Newtonian gravitational constant,
$f(R)$ is an arbitrary nonlinear function of $R$, and $S_m$ denotes the action
of matter.

For our purposes, we represent the function $f(R)$  in the form
\begin{equation}
\label{f_mod}
f(R)=R+\alpha h(R),
\end{equation}
where $h(R)$ is a new arbitrary function of $R$
and $\alpha$ is an arbitrary constant. When $\alpha=0$, one recovers Einstein's general relativity.
The corresponding field equations can be derived by
varying action \eqref{action_mod} with respect to the metric, yielding
\begin{equation}
\label{mod_Ein_eqs_gen}
\left(1+\alpha h_R\right) G_i^k-\frac{1}{2}\alpha\left(h-R\,h_R \right)\delta_i^k+
\alpha \left(\delta_i^k g^{m n}-\delta_i^m g^{k n}\right)\left(h_R\right)_{;m;n}=\frac{8\pi G}{c^4}T_i^k.
\end{equation}
Here $G_i^k\equiv R_i^k-\frac{1}{2}\delta_i^k R$ is the Einstein tensor, $h_R\equiv dh/dR$, and
the semicolon denotes the covariant derivative.

To obtain the modified Einstein equations and the equation for the fluid, we choose
the spherically symmetric metric in the form
\begin{equation}
\label{metric_schw}
ds^2=e^{\nu}(dx^0)^2-e^{\lambda}dr^2-r^2 \left(d\Theta^2+\sin^2\Theta\, d\phi^2\right),
\end{equation}
where $\nu$ and $\lambda$ are in general functions of  $r, x^0$,
and $x^0=c\, t$ is the time coordinate.

As a matter source in the field equations, we take an isotropic
fluid with
the energy-momentum tensor
\begin{equation}
\label{fluid_emt_anis}
T_{i}^k=\left(\varepsilon +p\right)u^k u_i-\delta_i^k p,
\end{equation}
where $\varepsilon$ is the fluid energy density and $p$ is the pressure.

The trace of Eq.~\eqref{mod_Ein_eqs_gen} yields the equation for the scalar curvature
\begin{equation}
\label{scal_cur_eq_gen}
-R+\alpha\left[h_R R-2 h+3\left(h_R\right)^{;i}_{;i}\right]=\frac{8\pi G}{c^4}T,
\end{equation}
where $T$ is the trace of the energy-momentum tensor \eqref{fluid_emt_anis}.

Using Eqs.~\eqref{metric_schw} and \eqref{fluid_emt_anis}, the $(^t_t)$, $(^r_r)$,   $(^\theta_\theta)$, and  $(^r_t)$ components of Eq.~\eqref{mod_Ein_eqs_gen}
can be written as
\begin{eqnarray}
\label{mod_00_gen}
&&
\left(1+\alpha h_R\right)
\left[-e^{-\lambda}\left(\frac{1}{r^2}-\frac{\lambda^\prime}{r}\right)+\frac{1}{r^2}\right]
-\frac{\alpha}{2}\left\{h-h_R R
+e^{-\lambda}\left[2 h_R^{\prime\prime}-\left(\lambda^\prime-\frac{4}{r}\right)h_R^{\prime}\right]-e^{-\nu}\dot{h}_R\dot\lambda
\right\}
=\frac{8\pi G}{c^4} \varepsilon,
 \\
\label{mod_11_gen}
&&\left(1+\alpha h_R\right)
\left[-e^{-\lambda}\left(\frac{1}{r^2}+\frac{\nu^\prime}{r}\right)+\frac{1}{r^2}\right]
-\frac{\alpha}{2}\left[h-h_R R-e^{-\nu}\left(2\ddot{h}_R-\dot{h}_R\dot\nu\right)
+e^{-\lambda}\left(\nu^\prime+\frac{4}{r}\right)h_R^{\prime}
\right]
=-\frac{8\pi G}{c^4} p,
\\
\label{mod_22_gen}
&&\left(1+\alpha h_R\right)
\left\{\frac{e^{-\lambda}}{2}\left[\frac{1}{r}\left(\lambda^\prime-\nu^\prime\right)+\frac{1}{2}\lambda^\prime\nu^\prime-\frac{1}{2}\nu^{\prime 2}-\nu^{\prime\prime}
\right]+\frac{e^{-\nu}}{2}\left(\ddot\lambda+\frac{1}{2}\dot\lambda^2-\frac{1}{2}\dot\lambda\dot\nu
\right)
\right\}\nonumber\\
&&-\frac{\alpha}{2}\left\{h-h_R R+e^{-\lambda}\left[2 h_R^{\prime\prime}-\left(\lambda^\prime-\nu^\prime-\frac{2}{r}\right)h_R^{\prime}\right]
-e^{-\nu}\left[2\ddot{h}_R+\left(\dot\lambda-\dot \nu\right)\dot{h}_R\right]
\right\}=-\frac{8\pi G}{c^4} p,
\\
\label{mod_10_gen}
&&-\left(1+\alpha h_R\right)\frac{e^{-\lambda}}{r}\dot\lambda-\alpha e^{-\lambda}
\left[\frac{1}{2}\left(\dot\lambda h_R^\prime+\nu^\prime \dot{h}_R\right)-\dot{h}_R^\prime
\right]=\frac{8\pi G}{c^4} \left(\varepsilon+p\right)u_0 u^1,
\end{eqnarray}
where the dot and prime denote differentiation with respect to $x^0$ and $r$, respectively.
In turn, Eq.~\eqref{scal_cur_eq_gen} yields
\begin{equation}
\label{scal_cur_eq_gen1}
-R+\alpha\left\{
-2 h+ h_R R+\frac{3}{2}\left[e^{-\nu}\left\{2\ddot{h}_R+\left(\dot\lambda-\dot\nu\right)\dot{h}_R\right\}-
e^{-\lambda}\left\{2 h_R^{\prime\prime}-\left(\lambda^\prime-\nu^\prime-\frac{4}{r}\right)h_R^\prime
\right\}
\right]
\right\}=\frac{8\pi G}{c^4}\left(\varepsilon-3 p\right).
\end{equation}

Finally, the $i=r$ component of the
law of conservation of energy and momentum, $T^k_{i;k}=0$, gives
\begin{equation}
\label{conserv_osc}
\frac{\partial T^0_1}{\partial x^0}+\frac{\partial T^1_1}{\partial r}+\frac{1}{2}\left(\dot{\nu}+\dot{\lambda}\right)T^0_1+
\frac{1}{2}\left(T_1^1-T_0^0\right)\nu^\prime+\frac{2}{r}\left[T_1^1-\frac{1}{2}\left(T^2_2+T^3_3\right)\right]=0.
\end{equation}

\section{Equilibrium configurations}
\label{stat_conf}

\subsection{Static equations}
General equations derived in the previous section can be employed to construct static solutions describing equilibrium configurations. For this purpose,
it is sufficient to set that all functions entering these equations depend on the radial coordinate $r$ only.
Also, bearing in mind the necessity of a physical interpretation of the results, it is convenient to introduce a new function $M(r)$, defined as
\begin{equation}
\label{metr_g11}
e^{-\lambda}=1-\frac{2 G M(r)}{c^2 r}.
\end{equation}
Then Eq.~\eqref{mod_00_gen} can be recast in the form
\begin{equation}
\label{mass_eq}
\left[1+\alpha\left(h_R+\frac{1}{2}r h_R^\prime\right)\right]\frac{d M}{d r}=
\frac{4\pi}{c^2} r^2 \varepsilon+\alpha \frac{c^2}{2 G}r^2
\left[\frac{1}{2}\left(h-h_R R\right)+h_R^{\prime}\left(\frac{2}{r}-\frac{3G M}{c^2 r^2}\right)+h_R^{\prime\prime}\left(1-\frac{2 G M}{c^2 r}\right)
\right].
\end{equation}

In GR (when $\alpha=0$),
the function $M(r)$ plays the role of the current mass inside a sphere of radius $r$. Then outside the fluid
 (i.e., when $\varepsilon=0$), $M=\text{const.}$ is the total gravitational mass of the object.
  A different situation occurs in the MGT (when $\alpha\neq 0$): outside the fluid
the scalar curvature is now nonzero (one can say that the star is surrounded by a gravitational sphere~\cite{Astashenok:2017dpo}).
This sphere gives an additional contribution to the total mass measured by a distant observer.
Depending on the sign of $\alpha$, the metric function $\lambda$ [and correspondingly the scalar curvature $R$ and
the mass function $M(r)$] either decays asymptotically or demonstrates an oscillating behavior. In the latter case $M(r)$ cannot already be regarded as the mass function.
Consistent with this, here, we use only such $\alpha$'s that ensure a nonoscillating behavior of $M(r)$; this enables us to interpret $M(r\to \infty)$ as the total
mass.

In turn, the conservation law given by Eq.~\eqref{conserv_osc} yields the  equation
\begin{equation}
\label{conserv_gen}
\frac{d p}{d r}=-\frac{1}{2}\left(\varepsilon+p\right)\frac{d \nu}{d r}.
\end{equation}

For a complete description of the  configuration under consideration, the above equations
must be supplemented by an equation of state (EoS) for the fluid.
Here, for the sake of simplicity, we consider
a barotropic EoS where the pressure is a function of the mass density $\rho_b$.
For our purpose,  we restrict ourselves to a simplified variant of the EoS,
where a more or less realistic matter
EoS is approximated in the form of the following polytropic EoS:
\begin{equation}
\label{eqs_NS_WH}
p=K \rho_{b}^{1+1/n}, \quad \varepsilon = \rho_b c^2 +n p,
\end{equation}
with the constant $K=k c^2 (n_{b}^{(ch)} m_b)^{1-\gamma}$,
the polytropic index $n=1/(\gamma-1)$,
and $\rho_b=n_{b} m_b$ denotes the rest-mass density
of the fluid. Here, $n_{b}$ is the baryon number density,
$n_{b}^{(ch)}$ is a characteristic value of $n_{b}$,
$m_b$ is the baryon mass,
and $k$ and $\gamma$ are parameters
whose values depend on the properties of the matter.

Next, introducing the new variable $\theta$,
$$\rho_b=\rho_{b c} \theta^n,$$
where $\rho_{b c}$ is the central density of the fluid,
we may rewrite the pressure and the energy density, given by
Eq.~\eqref{eqs_NS_WH}, in the form
$$p=K\rho_{b c}^{1+1/n} \theta^{n+1}, \quad
\varepsilon =  \left( \rho_{b c} c^2 +
  n K \rho_{b c}^{1 + {1}/{n} } \theta \right) \theta^n.$$
Making use of these expressions, from Eq.~\eqref{conserv_gen},
we obtain
for the internal region with $\theta \ne 0$,
\begin{equation}
\label{conserv_3}
2\sigma(n+1)\frac{d\theta}{d r}=
-\left[1+\sigma(n+1) \theta\right]\frac{d\nu}{dr},
\end{equation}
where $\sigma=K \rho_{b c}^{1/n}/c^2=p_c/(\rho_{b c} c^2)$ is a relativity parameter,
related to the central pressure $p_c$ of the fluid.
This equation may be integrated to give the metric function $e^{\nu}$
in terms of $\theta$,
$$e^{\nu}=e^{\nu_c}\left[\frac{1+\sigma (n+1)}{1+\sigma (n+1)\theta}\right]^{2},$$
and $e^{\nu_c}$ is the value of $e^{\nu}$ at the center where $\theta=1$.
The integration constant $\nu_c$ is fixed
by the requirement of the asymptotical flatness of the spacetime,
i.e., $e^{\nu}=1$ at infinity.

\subsection{Numerical results}

Thus, we have four unknown functions~-- $R, \theta$, $\nu$, and $M$~-- for which there are four equations,
\eqref{mod_22_gen}, \eqref{scal_cur_eq_gen1}, \eqref{mass_eq}, and \eqref{conserv_3}
whose solution will depend on the choice of the particular type of gravity theory, i.e., of the function $h$.

In the present paper, we consider the simplest case of quadratic gravity when $h=R^2$,
which is often discussed in the literature as a
viable alternative cosmological model describing the accelerated expansion of the early and present Universe~\cite{Nojiri:2010wj,Nojiri:2017ncd}.
For such gravity theory,
the value of the free parameter  $\alpha$ appearing in \eqref{f_mod} is constrained from observations as follows:
(i)~in the weak-field limit, it is constrained by binary pulsar data as $|\alpha| \lesssim 5\times 10^{15} \text{cm}^2$~\cite{Naf:2010zy};
 (ii)~in the strong gravity regime, the constraint is $|\alpha| \lesssim 10^{10} \text{cm}^2$~\cite{Arapoglu:2010rz}.
 Consistent with this, for the calculations presented below, we
take $\alpha = -10^{10} \text{cm}^2$ (notice that we take an opposite sign for $\alpha$
as compared with that used in Ref.~\cite{Astashenok:2017dpo} since here we employ another metric signature).
If one takes another sign of $\alpha$,
it can result in the appearance of  ghost modes and
instabilities in the cosmological context~\cite{Barrow:1983rx};
also, in this case, the scalar curvature  $R$ demonstrates an oscillating behavior outside the star,
which appears to be unacceptable if one intends to
construct realistic models of compact configurations (for a detailed discussion, see Ref.~\cite{Astashenok:2017dpo}).

For numerical calculations, it is convenient to rewrite the equations in terms of the dimensionless variables
\begin{equation}
\label{dmls_var}
x=r/L, \quad v(x)=\frac{M(r)}{4\pi \rho_{bc} L^3}, \quad \Sigma=R L^2, \quad \bar{\alpha}=\alpha/L^2,\quad
\text{where}
\quad L=\frac{c}{\sqrt{8\pi G \rho_{bc}}}.
\end{equation}
As a result, we get the static equations
\begin{eqnarray}
\label{mod_Einstein-00_stat}
&&v^\prime=\frac{x^2}{1+\bar\alpha\left(2 \Sigma+x \Sigma^\prime\right)}\left\{
\left(1+\sigma n \theta\right)\theta^n+\frac{\bar\alpha}{2}\left[
- \Sigma^2+2 \frac{\Sigma^\prime}{x}\left(4-3\frac{v}{x}\right)+4  \Sigma^{\prime\prime}\left(1-\frac{v}{x}\right)
\right]
\right\},
 \\
\label{mod_Einstein-22_stat}
&&-\left(1-\frac{v}{x}\right)\left(\nu^{\prime\prime}+\frac{\nu^{\prime 2}}{2}\right)
+\frac{v^\prime}{x^2}+\left(v^\prime+\frac{v}{x}-2\right)\frac{\nu^\prime}{2 x}
-\frac{v}{x^3}+2\sigma\theta^{n+1}\nonumber \\
&&+\bar\alpha\Big\{
\Sigma^2+2\Sigma^\prime\left[-\frac{2}{x}+\frac{v}{x^2}+\frac{v^\prime}{x}-\nu^\prime\left(1-\frac{v}{x}\right)
\right]-4\left(1-\frac{v}{x}\right)\Sigma^{\prime\prime}+\frac{v \Sigma}{x}\left[2\nu^{\prime\prime}+\nu^{\prime 2}+\frac{\nu^\prime}{x}-\frac{2}{x^2}
\right] \nonumber\\
&&+\Sigma\left[v^\prime\left(\frac{\nu^\prime}{x}+\frac{2}{x^2}\right)-2\frac{\nu^\prime}{x}-2\nu^{\prime\prime}-\nu^{\prime 2}
\right]
\Big\}=0,
\\
\label{conserv_stat}
&&2\sigma(n+1)\theta^\prime=-\left[1+\sigma\left(n+1\right)\theta\right]\nu^\prime,
\\
\label{curv_stat}
&&\Sigma+\left[1+\sigma(n-3)\theta\right]\theta^n+\bar\alpha\Big\{6\left(1-\frac{v}{x}\right)\Sigma^{\prime\prime}
+3\frac{\Sigma^\prime}{x}\left[-\frac{v}{x}\left(3+x \nu^\prime\right)+4-v^\prime+x \nu^\prime
\right]
\Big\}=0,
\end{eqnarray}
which follow from Eqs.~\eqref{mass_eq}, \eqref{mod_22_gen}, \eqref{conserv_3}, and \eqref{scal_cur_eq_gen1}, respectively.
When $\bar\alpha=0$, one recovers the general-relativity equations.

It is also convenient to recast the mass and radius of the configuration in terms of the parameters $K, n$, and $\sigma$~\cite{Tooper2}.
By eliminating $\rho_{b c}$ from the expressions for $x$ and $v$ in Eq.~\eqref{dmls_var}, we obtain
$$r=r^*\sigma^{-n/2} x, \quad M(r)=M^*\sigma^{-n/2}v(x), \quad \alpha=\alpha^* \sigma^{-n}\bar\alpha,$$
where $r^*=(8\pi G)^{-1/2}K^{n/2}c^{1-n},
M^*=(1/4)(2\pi)^{-1/2}G^{-3/2}K^{n/2}c^{3-n}$, and $\alpha^*=(8\pi G)^{-1}K^{n}c^{2(1-n)}$.
The quantities $r^*$ and $M^*$ define the scales of the radius and mass.

Equations \eqref{mod_Einstein-00_stat}-\eqref{curv_stat} are to be solved subject to the boundary conditions given in the neighborhood of the center by the expansions
\begin{equation}
\label{bound_mod_Ein}
\theta\approx 1+\frac{1}{2}\theta_2 x^2, \quad \nu\approx\nu_c+\frac{1}{2}\nu_2 x^2, \quad v\approx \frac{1}{6} v_3 x^3, \quad
\Sigma\approx\Sigma_c+\frac{1}{2}\Sigma_2 x^2,
\end{equation}
where the expansion coefficients $\theta_2, \nu_2, v_3,$ and $\Sigma_2$
are determined from Eqs.~\eqref{mod_Einstein-00_stat}-\eqref{curv_stat}. The central value of the scalar curvature $\Sigma_c$
is an eigenparameter of the problem, and it
is chosen so that asymptotically $\Sigma(x\to \infty)\to 0$.

The integration of Eqs.~\eqref{mod_Einstein-00_stat}-\eqref{curv_stat}
is performed numerically from the center (i.e., from $x\approx 0$) to the point $x=x_b$,
where the fluid density goes to zero.
 We take this point to be a boundary of the star.
In turn, for $x>x_b$ the matter is absent, i.e., $\rho_b=p=0$. In GR,
this corresponds to the fact that the scalar curvature $\Sigma=0$.
But in the MGT this is not the case: there is an external
 gravitational sphere around the star where $\Sigma\neq 0$.
Consistent with this, the internal solutions should be matched with the external
ones at the edge of the fluid. This is done by equating the
corresponding values of both the scalar curvature and the metric functions.

For negative $\alpha$'s employed here,
the scalar curvature is  damped exponentially fast outside the fluid as
$
\Sigma\sim \exp{\left(-x/\sqrt{6|\bar \alpha|}\right)}/ x.
$
This enables us to introduce a  well-defined notion for the Arnowitt-Desser-Misner mass through Eq.~\eqref{metr_g11},
unlike the case of  positive $\alpha$'s for which $\Sigma$ demonstrates an oscillating behavior~\cite{Astashenok:2017dpo}.

\begin{figure}[t]
\centering
  \includegraphics[height=7cm]{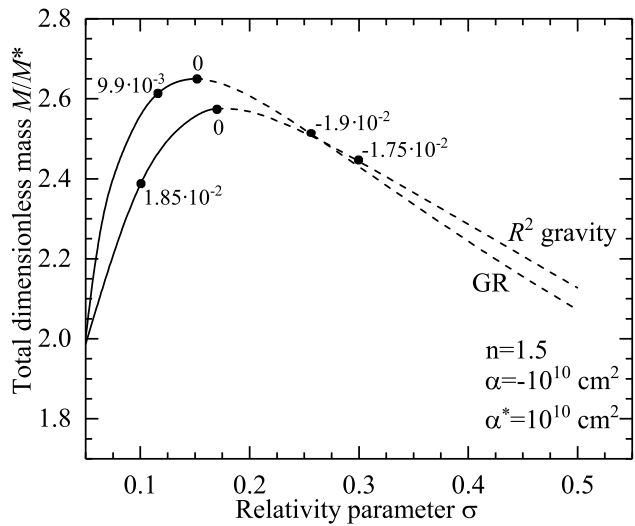}
\caption{The dimensionless total mass versus the relativity parameter $\sigma$.  The numbers near the curves denote the values of the square of the lowest eigenfrequency
$\bar \omega^2$ corresponding to the configuration at a given point of the curve.
The segments of the curves corresponding to stable configurations are shown as solid lines, whereas the unstable segments are shown dashed.
}
\label{fig_M_sigma}
\end{figure}

The results of numerical calculations are shown in Fig.~\ref{fig_M_sigma}, where the dependence of the total mass on
the relativity parameter $\sigma$ (or the central density $\rho_{bc}$) is plotted. It is seen that both in GR and in the MGT the curves
have a maximum.
In GR, such a maximum corresponds
to the transition from stable to unstable systems, and this is confirmed by the linear stability analysis~\cite{Tooper2}.
In the next section, we will study this problem for the case of  $R^2$ gravity under consideration.

\section{Linear stability analysis}
\label{stab_anal}

Consider that the equilibrium systems described above are perturbed in such a way that spherical
symmetry is maintained. In obtaining the equations for the perturbations, we will neglect all quantities which are
of the second and higher order. The components of the four-velocity in the metric~\eqref{metric_schw} are given by \cite{Chandrasekhar:1964zz}
$$
u^0=e^{-\nu_0/2}, \quad u_0=e^{\nu_0/2}, \quad u^1=e^{-\nu_0/2} \mathpzc{v}, \quad u_1=-e^{\lambda_0-\nu_0/2} \mathpzc{v},
$$
with the three-velocity $\mathpzc{v}=d r/d x^0 \ll 1$.
The index 0 in the metric functions indicates the static, zeroth order solution of the gravitational equations.

Now we consider perturbations of the static solutions of the form
\begin{equation}
\label{perturbations}
y=y_0+y_p ~,
\end{equation}
where the index 0 refers to the static solutions, the index $p$ indicates the perturbation,
and $y$ denotes one of the functions $\lambda, \nu, \varepsilon, p$ or the scalar curvature $R$.
We will use the variational approach of Ref.~\cite{Chandrasekhar:1964zz} when one introduces a ``Lagrangian displacement'' $\zeta$ with respect to $x^0$,
$\mathpzc{v}=\partial \zeta/\partial x^0$.
Then, substituting the expansions \eqref{perturbations} in Eqs.~\eqref{mod_00_gen}-\eqref{conserv_osc} and seeking solutions in a
harmonic form
$$y_p(x^0,r) = \tilde{y}_p(r) e^{i\omega x^0}$$
[for convenience, we hereafter drop the tilde sign on $\tilde{y}_p(r)$],
one can obtain the following set of equations for the perturbations $\theta_p, \Sigma_p$, and $\lambda_p$:
\begin{eqnarray}
\label{eq_theta_pert}
&&\sigma(n+1) s_1 \theta_0^n \theta_p^\prime+
\frac{1}{2}\theta_0^{n}\Big\{\sigma(n+1)x\theta_0^n e^{\lambda_0}\left[1+\sigma(n+1)\theta_0\right]+
s_1\left[\sigma(n+1)^2\nu_0^\prime+\frac{n}{\theta_0}\Big(2\sigma (n+1)\theta_0^\prime+\nu_0^\prime\Big)\right]
\Big\}\theta_p\nonumber\\
&&+\frac{\lambda_p}{2x}e^{-\nu_0}\Big\{
8\bar\alpha^2\bar\omega^2 \Sigma_0^2+2\bar\omega^2\left(1+\bar\alpha x\Sigma_0^\prime\right)^2+
2\bar\alpha \Sigma_0\left[4\bar\omega^2\left(1+\bar\alpha x\Sigma_0^\prime\right)+e^{\nu_0}\theta_0^n\left(1+x\nu_0^\prime\right) s_2
\right]\nonumber \\
&&+e^{\nu_0}\theta_0^n\left[1+x\nu_0^\prime+\bar\alpha x\Sigma_0^\prime\left(4+x \nu_0^\prime\right)
\right]s_2
\Big\}-\frac{\bar\alpha}{2}e^{-\nu_0}\left[8\bar\alpha\bar\omega^2\Sigma_0+4\bar\omega^2\left(1+\bar\alpha x\Sigma_0^\prime\right)+e^{\nu_0}\theta_0^n\left(4+x\nu_0^\prime\right) s_2
\right]\Sigma_p^\prime
\nonumber\\
&&+\frac{\bar\alpha}{2 x}e^{-\nu_0}\left\{
2\bar\omega^2 x s_1 \nu_0^\prime+2 s_2 e^{\nu_0} \theta_0^n \left[
e^{\lambda_0}\left(1-\bar\omega^2 x^2 e^{-\nu_0}+\frac{1}{2}x^2 \Sigma_0\right)-x\nu_0^\prime-1
\right]
\right\}\Sigma_p=0,
\end{eqnarray}
\begin{eqnarray}
\label{eq_R_pert}
&&\bar\alpha s_1 \Sigma_p^{\prime\prime}-\frac{\bar\alpha}{2x}\left\{
x\left(1+\bar\alpha x \Sigma_0^\prime\right)\lambda_0^\prime-x\nu_0^\prime+2\bar\alpha \Sigma_0\left[x\left(\lambda_0^\prime-\nu_0^\prime\right)-4\right]-4
\right\}\Sigma_p^\prime-\frac{\bar\alpha}{2}\Sigma_0^\prime s_1 \lambda_p^\prime \nonumber\\
&&+\frac{e^{-\nu_0}}{6x}\left\{
x e^{\lambda_0}\left(e^{\nu_0}+6\bar\alpha \bar\omega^2\right)+\bar\alpha x e^{\lambda_0}\Sigma_0\left[
2\left(e^{\nu_0}+6\bar\alpha\bar\omega^2\right)+3 \bar\alpha x e^{\nu_0}\Sigma_0^\prime
\right]+\bar\alpha e^{\nu_0}\Sigma_0^\prime\left[
-6\bar\alpha\left(1+x\nu_0^\prime\right)+e^{\lambda_0}\left(x^2+6\bar\alpha\right)
\right]
\right\}\Sigma_p\nonumber\\
&&+\frac{e^{\lambda_0}}{6}\theta_0^{n-1}\left\{
n\left(1+\bar\alpha x \Sigma_0^\prime\right)+2\bar\alpha \left[n+\sigma\left(n^2-2n-3\right)\theta_0
\right]\Sigma_0+\sigma(n+1)\theta_0\left[n-3+\bar\alpha n x\Sigma_0^\prime\right]
\right\}\theta_p\nonumber\\
&&+\frac{\bar\alpha}{2x}\left\{
\bar\alpha x^2\Sigma_0^{\prime 2}\lambda_0^\prime-2x\left(1+2\bar\alpha \Sigma_0\right)\Sigma_0^{\prime\prime}+
\Sigma_0^\prime\left[\left(x\lambda_0^\prime-3\right)\left(1+2\bar\alpha\Sigma_0\right)-2\bar\alpha x^2\Sigma_0^{\prime\prime}
\right]
\right\}\lambda_p=0,
\end{eqnarray}
\begin{eqnarray}
\label{eq_lambda_pert}
&&\bar\alpha \Sigma_p^{\prime\prime}+\frac{\bar\alpha}{x}\left(2-\frac{1}{2}x\lambda_0^\prime\right)\Sigma_p^\prime-
\bar\alpha\left(\frac{1}{2}e^{\lambda_0}\Sigma_0+\frac{x\lambda_0^\prime+e^{\lambda_0}-1}{x^2}\right)\Sigma_p-\frac{s_1}{2x}\lambda_p^\prime \nonumber\\
&&+\frac{1}{2x^2}\left[\left(x\lambda_0^\prime-1\right)\left(1+2\bar\alpha\Sigma_0\right)+\bar\alpha x \Sigma_0^\prime\left(x\lambda_0^\prime-4\right)-2\bar\alpha x^2\Sigma_0^{\prime\prime}
\right]\lambda_p+\frac{n}{2}e^{\lambda_0}\theta_0^{n-1}s_2\theta_p=0,
\end{eqnarray}
where $s_1=1+\alpha\left(2 \Sigma_0+x \Sigma_0^\prime\right), s_2=1+\sigma(n+1)\theta_0$, and the dimensionless frequency $\bar\omega=L \omega$.
Here, Eq.~\eqref{eq_theta_pert} follows from the conservation law~\eqref{conserv_osc}, Eq.~\eqref{eq_R_pert} follows from the equation for the scalar curvature~\eqref{scal_cur_eq_gen1},
Eq.~\eqref{eq_lambda_pert} is the $(^t_t)$ component~\eqref{mod_00_gen}.
In deriving  Eq.~\eqref{eq_theta_pert}, we have used the expression (here $\psi=\zeta/L$ is the dimensionless Lagrangian displacement)
$$
\lambda_p=-\frac{x}{1+\bar \alpha \left(2 \Sigma_0+x \Sigma_0^\prime\right)}\left\{
e^{\lambda_0}\theta_0^n\left[1+\sigma (n+1)\theta_0\right]\psi+\bar\alpha\left(\Sigma_p\nu_0^\prime-2\Sigma_p^\prime\right)
\right\}
$$
[which follows from the $(^r_t)$ component \eqref{mod_10_gen}],
expressing from it $\psi$ and eliminating it from~\eqref{eq_theta_pert}.

For this set of equations, we choose
the following boundary conditions near the center $x=0$:
\begin{equation}
\label{bound_cond_pert}
\theta_p\approx \theta_{pc}+\frac{1}{2}\theta_{p2} x^2, \quad \Sigma_p\approx \Sigma_{pc}+\frac{1}{2}\Sigma_{p2} x^2, \quad
\lambda_p\approx \frac{1}{2}\lambda_{p2} x^2,
\end{equation}
where the expansion coefficients $\theta_{pc}$ and $\Sigma_{pc}$ are arbitrary and
$\theta_{p2} , \lambda_{p2}$, and $\Sigma_{p2}$ can be found from Eqs.~\eqref{eq_theta_pert}-\eqref{eq_lambda_pert}.

The set of equations~\eqref{eq_theta_pert}-\eqref{eq_lambda_pert}
together with the boundary conditions \eqref{bound_cond_pert}
defines an eigenvalue problem for $\bar\omega^2$.
The question of stability is therefore reduced to a study of the possible
values of $\bar\omega^2$.
If any of the values of $\bar\omega^2$ are found to be negative,
then the perturbations will increase and the
configurations in question will be unstable against radial oscillations.

The choice of eigenvalues of $\bar \omega^2$ is carried out such that we have asymptotically decaying solutions for the perturbations $\Sigma_p$ and $\lambda_p$.
In doing so, it is necessary to ensure the following properties of the solutions:
(i)~The function  $\theta_p$ must be finite (though not necessarily zero) at the boundary of the star.
This is sufficient to ensure that the perturbation of the fluid pressure $p_p\sim \theta_0^n \theta_p$ meets the condition $p_p=0$
at the edge of the star where $\theta_0= 0$ [see, e.g., Eq.~(60) in Ref.~\cite{Chandrasekhar:1964zz}].
 (ii)~The function $\lambda_p$ must be nodeless; this corresponds to a zero mode of the solution
 (on this point, see, e.g., Ref.~\cite{Gleiser:1988ih}).

In this connection,
it is useful to write out the asymptotic behavior of the solutions.

\noindent (A) {\it Static solutions}:
$$v \to v_{\infty}-C_{\Sigma_0} \sqrt{\frac{2 |\bar \alpha|}{3}}\,x \exp{\left(-x/\sqrt{6|\bar \alpha|}\right)}, \quad
\Sigma_0 \to -C_{\Sigma_0} \exp{\left(-x/\sqrt{6|\bar \alpha|}\right)}\Big/x, \quad
e^{\nu_0}\to 1-v_{\infty}/x.$$

\noindent (B) {\it Perturbations}:
$$\Sigma_p\to C_{\Sigma_p} \exp{\left(-\sqrt{\frac{1+6|\bar \alpha|\bar\omega^2}{6|\bar \alpha|}}\,x\right)}\Big/x, \quad
\lambda_p \to v_{\infty}/x.$$
Here, $v_{\infty}$ is an asymptotic value of the mass function $v$,
  $C_{\Sigma_0}>0$ and $C_{\Sigma_p}$ are integration constants.





\begin{figure}[t]
\centering
  \includegraphics[height=4cm]{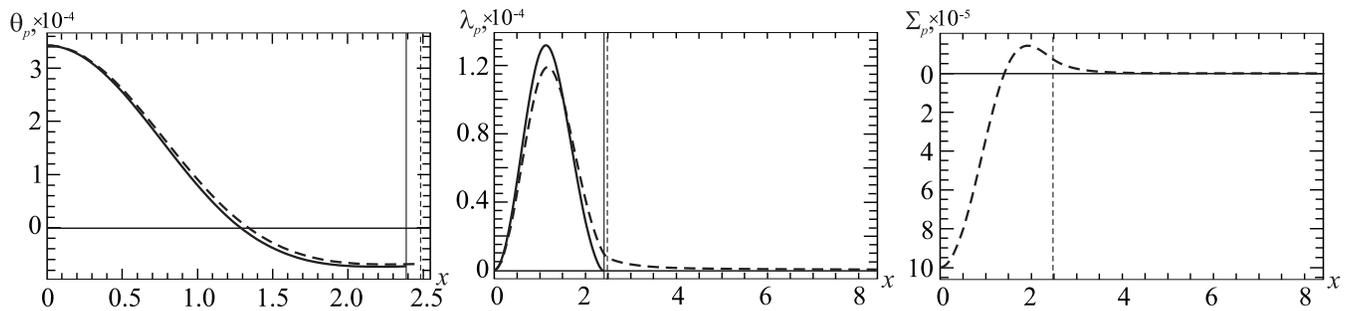}
\caption{The typical behavior of the perturbations within GR (solid curves) and $R^2$ gravity (dashed curves).
The graphs are plotted for the case of $\sigma\approx 0.151$ (GR) and of $\sigma= 0.17$ ($R^2$ gravity)
that correspond to the maxima of the mass curves (cf. Fig.~\ref{fig_M_sigma}).
The thin vertical lines denote the boundaries of the fluid $x=x_b$.
}
\label{fig_pertubs}
\end{figure}

\begin{table}[h!]
 \caption{The computed values of the square of the lowest eigenfrequency
   $\bar \omega^2$ and of the eigenparameter $\theta_{pc}$ for the configurations within $R^2$ gravity. For all the cases $\Sigma_{pc}=-10^{-4}$.
 The eigenparameter $\Sigma_c$ is the central value of the scalar curvature from Eq.~\eqref{bound_mod_Ein}.}
\vspace{.3cm}
\begin{tabular}{|c|c|c|c|}
\hline
$\sigma$ &$\Sigma_c$ & $\bar \omega^2$ & $\theta_{pc}$\\
\hline
0.1&-0.494418253625 & 0.0185 & 0.00020166833\\
\hline
0.17&-0.3646950715 & $\approx 0$ & 0.00034320286\\
\hline
0.25&-0.26573801601 &-0.0118 & 0.000644452\\
\hline
0.3&-0.2202699654 & -0.0175 & 0.000993096\\
\hline
\end{tabular}
\label{tab1}
\end{table}

Examples of solutions for the perturbations $\theta_p, \Sigma_p$, and $\lambda_p$ are shown in Fig.~\ref{fig_pertubs}.
The procedure for determining the eigenvalues of $\bar \omega^2$ is as follows:
\begin{enumerate}
\itemsep=-0.2pt
\item[(1)]
 In the case of GR (when $\bar\alpha=0$), there are only two equations  \eqref{eq_theta_pert} and \eqref{eq_lambda_pert}
 for the functions  $\theta_p$ and $\lambda_p$. Considering that these equations are linear, the rescaling of the central
$\theta_{pc}\to \beta \theta_{pc}$ results in the corresponding rescaling of the metric perturbation
 $\lambda_p\to \beta \lambda_p$, but the qualitative behavior of the solutions remains unchanged.  That is, it is possible to take any central
 $\theta_{pc}\ll 1$, and the eigenfrequency $\bar \omega^2$ will not change.

\item[(2)]  In the case of $R^2$ gravity (when $\bar\alpha\neq0$),
all three equations~\eqref{eq_theta_pert}-\eqref{eq_lambda_pert} need to be solved. In doing so, there are two arbitrary central values
 $\theta_{pc}$ and $\Sigma_{pc}$.
 Numerical calculations indicate that to ensure regular asymptotically decaying solutions for the perturbations
$\Sigma_p$ and $\lambda_p$ one has to adjust the values both of the eigenfrequency
$\bar \omega^2$ and of one of these two arbitrary parameters. That is, either $\theta_{pc}$ or $\Sigma_{pc}$ is an eigenparameter of the problem.
The corresponding numerical values of these parameters are given in Table~\ref{tab1} for several values of $\sigma$.
It is seen from the table and Fig.~\ref{fig_M_sigma} that the square of the lowest eigenfrequency $\bar \omega^2$
is positive to the left of the maximum and negative to the right of it. That is, as in GR,
in $R^2$  gravity under consideration the transition from stable to unstable systems occurs strictly at the maximum of the mass.
\end{enumerate}

Summarizing the results obtained,
within $R^2$ gravity,  we have examined the question of stability of compact configurations supported by a polytropic fluid against linear radial perturbations. In contrast to the
studies performed earlier in the literature, here the calculations have been carried out in the Jordan frame to avoid the potential difficulties related to the conformal transformation
to the Einstein frame (see Introduction). In doing so, we regard the scalar curvature as a dynamical variable for which the behavior of the corresponding perturbation modes is studied as well.
As a result, it is shown that, as in GR, within the framework of $R^2$ gravity the transition from stable to unstable configurations takes place at the point of the maximum of the curve mass-central density of the fluid.
One may expect that similar results will also be obtained for another, more realistic EoSs of matter,
including those that are used in constructing models of neutron stars (see, e.g., Refs.~\cite{Astashenok:2017dpo,Folomeev:2018ioy,Astashenok:2020isy}).

\section*{Acknowledgements}

The authors are very grateful to S.~Odintsov for fruitful discussions and comments.
We gratefully acknowledge support provided by Grant No.~BR05236322
in Fundamental Research in Natural Sciences by the Ministry of Education and Science of the Republic of Kazakhstan.
We are also grateful to the Research Group Linkage Programme of the Alexander von Humboldt Foundation for the support of this research.

\end{document}